\documentclass[12pt,twoside]{article}
\usepackage{fleqn,espcrc1}
\usepackage{hyperref}

\usepackage{graphicx}

\title{The Charge Form Factor of the Neutron from 
$^2\vec{\rm H}$($\vec{e}$, $e^\prime$ $n$)$p$}

\begin{document}

\maketitle %
\setlength{\parindent}{0pt}%
\href{email:igorp@nikhef.nl}{I.~Passchier}$\rm^{a}$, L.~D.~van~Buuren$\rm^{a,b}$
D.~Szczerba$\rm^{c}$, R.~Alarcon$\rm^{d}$, Th.~S.~Bauer$\rm^{a,e}$,
D.~Boersma$\rm^{a,e}$, J.~F.~J.~van~den~Brand$\rm^{a,b}$,
H.~J.~Bulten$\rm^{a,b}$, M.~Ferro-Luzzi$\rm^{a,b}$,
D.~W.~Higinbotham$\rm^{f}$, C.~W.~de~Jager$\rm^{a,g}$,
S.~Klous$\rm^{a,b}$, H.~Kolster$\rm^{a,b}$, J.~Lang$\rm^c$,
D.~Nikolenko$\rm^h$, G.~J.~Nooren$\rm^a$, B.~E.~Norum$\rm^f$,
H.~R.~Poolman$\rm^{a,b}$, I.~Rachek$\rm^h$, M.~C.~Simani$\rm^{a,b}$,
E.~Six$\rm^{d}$, H.~de~Vries$\rm^a$, K.~Wang$\rm^f$, Z.-L.~Zhou$\rm^{i}$\\

$\rm^{a}$\href{http://www.nikhef.nl}{NIKHEF}, NL-1009 DB Amsterdam, The Netherlands\\
$\rm^{b}$Dept. of Physics and Astronomy, VU, NL-1081 HV, Amsterdam, 
The Netherlands\\
$\rm^{c}$Institut f\" ur Teilchenphysik, ETH, CH-8093 Z\" urich, Switzerland\\
$\rm^{d}$Dept. of Physics and Astronomy, ASU, Tempe, AZ 85287, USA\\
$\rm^{e}$Physics Dept., UU, NL-3508 TA Utrecht, The Netherlands\\
$\rm^{f}$Dept. of Physics, UVa, Charlottesville, VA 22901, USA\\
$\rm^{g}$TJNAF, Newport News, VA 23606, USA\\
$\rm^{h}$BINP, Novosibirsk, 630090, Russian Federation\\
$\rm^{i}$Laboratory for Nuclear Science, MIT, Cambridge, MA 02139, USA\\
\setlength{\parindent}{1em}

\begin{abstract}
  We report on the first measurement of spin-correlation parameters in
  quasifree electron scattering from vector-polarized deuterium.
  Polarized electrons were injected into an electron storage ring at a
  beam energy of 720~MeV.  A Siberian snake was employed to preserve
  longitudinal polarization at the interaction point.
  Vector-polarized deuterium was produced by an atomic beam source and
  injected into an open-ended cylindrical cell, internal to the
  electron storage ring. The spin correlation parameter $A^V_{ed}$ was
  measured for the reaction $^2 \vec{\rm H}(\vec e,e^\prime n)p$ at a
  four-momentum transfer squared of 0.21~(GeV/$c$)$^2$ from which a
  value for the charge form factor of the neutron was extracted.
\end{abstract}

\section{Motivation}
The charge distribution of the neutron is described by the charge
form factor $G_E^n$, which is related to the Fourier transform of the
distribution and is generally expressed as a function of $Q^2$, the
square of the four-momentum transfer.  Data on $G_E^n$ are important
for our understanding of the nucleon and are essential for the
interpretation of electromagnetic multipoles of nuclei, e.g. the
deuteron.

Since a practical target of free neutrons is not available,
experimentalists mostly resorted to (quasi)elastic scattering of
electrons from unpolarized deuterium\cite{Lun93,Platchkov} to
determine this form factor.  The shape of $G_E^n$ as function of $Q^2$
is relatively well known from high precision elastic electron-deuteron
scattering\cite{Platchkov}, but the absolute scale still contains a
systematic uncertainty of about 50\%. The slope of $G_E^n$ at
$Q^2=0$~(GeV/c)$^2$ is known from measurements where thermal neutrons
are scattered from atomic electrons ~\cite{Kop97}.

The systematic uncertainties can be significantly reduced
through the measurement of electronuclear spin observables.  The
scattering cross section with both longitudinal polarized electrons
and a polarized target for the $^2 \vec{\rm H}(\vec e,e^\prime
N)$ reaction, can be written as~\cite{Arenh}
\begin{equation}
  S = S_0 \left\{ 1 + P_1^d A^V_d + P_2^d A^T_d + h( A_e + P_1^d
    A^V_{ed} + P_2^d A^T_{ed}) \right\},
\end{equation}
where $S_0$ is the unpolarized cross section, $h$ the polarization of
the electrons, and $P_1^d$ ($P_2^d$) the vector (tensor) polarization
of the target. The target analyzing powers and spin-correlation
parameters ($A_i$), depend on the orientation of the nuclear spin.
The polarization direction of the deuteron is defined by the angles
$\Theta_d$ and $\Phi_d$ in the frame where the $z$-axis is along the
direction of the three-momentum transfer ($\bf{q}$) and the $y$-axis
is defined by the vector product of the incoming and outgoing electron
momenta.  The observable $A^V_{ed}(\Theta_d=90^\circ,\Phi_d=0^\circ)$
contains an interference term, where the effect of the small charge
form factor is amplified by the dominant magnetic form factor (see
e.g.  Refs.~\cite{Arenh,Bill}).  In the present paper we describe a
measurement performed at NIKHEF (Amsterdam), which uses a stored
polarized electron beam and a vector-polarized deuterium target, to
determine $G_E^n$ via a measurement of
$A^V_{ed}(90^\circ,0^\circ)$.

\section{Experimental setup}
The experiment was performed with a polarized gas target internal to
the AmPS electron storage ring.  An atomic beam source
(ABS)\cite{ABSRF,Buuren} was used to inject a flux of $4.6\times
10^{16}$ deuterium atoms/s into a cooled storage cell.

Polarized electrons were produced by photo-emission from a
strained-layer semiconductor cathode (InGaAsP)\cite{PES}. After linear
acceleration to 720~MeV the electrons were injected and stacked in the
AmPS storage ring.  Every 5 minutes the ring was refilled, after
reversal of the electron polarization at the source.  The polarization
of the stored electrons was maintained by setting the spin tune to 0.5
with a strong solenoidal field (using the {Siberian snake}
principle\cite{snake}).


Scattered electrons were detected in the large-acceptance magnetic
spectrometer~\cite{BB}.  The electron detector was positioned
at a central angle of 40$^\circ$, resulting in a central value of $Q^2
= 0.21$(GeV/c)$^2$.  Neutrons and protons were detected in a
time-of-flight (TOF) system made of two subsequent and identical
scintillator arrays.  Each of the four bars in an array was preceded
by two plastic scintillators used to identify and/or veto charged
particles.  By simultaneously detecting protons and neutrons in the
same detector, one can construct asymmetry ratios for the two reaction
channels $^2\vec{\rm H}(\vec e,e^\prime p)n$ and $^2\vec{\rm H}(\vec
e,e^\prime n)p$, in this way minimizing systematic uncertainties
associated with the deuteron ground-state wave function, absolute beam
and target polarizations, and possible dilution by cell-wall
background events.

\section{Results}
An experimental asymmetry ($A_{exp}=\frac{N_+ - N_-}{N_++N_-}$) can be constructed,
where $N_\pm$ is the number of events that pass the selection
criteria, with $h P_1^d$ either positive or negative. $A_{exp}$ for
the $^2\vec{\rm H}(\vec e,e^\prime p)n$-channel, integrated up to a
missing momentum of 200~MeV/$c$, was used to
determine the effective product of beam and target polarization by
comparing to the predictions of the model of Arenh\"ovel \emph{et
  al.}~\cite{Arenh}.  This advanced, non-relativistic model has shown to
provide good descriptions for quasifree proton knockout from
tensor-polarized deuterium\cite{eep}.  Finite acceptance effects were
taken into account with a Monte Carlo code. 

The spin-correlation parameter for the neutron events was obtained
from the experimental asymmetry by correcting for the contribution of
protons misidentified as neutrons (less than 1\%, as determined from a
calibration with the reaction $^1$H($e,e^\prime p$)), and for the
product of beam and target polarization, as determined from the
$^2\vec{\rm H}(\vec e,e^\prime p)n$ channel.

\begin{figure}[htb]
  \begin{minipage}[t]{75mm}
    \includegraphics[width=75mm]{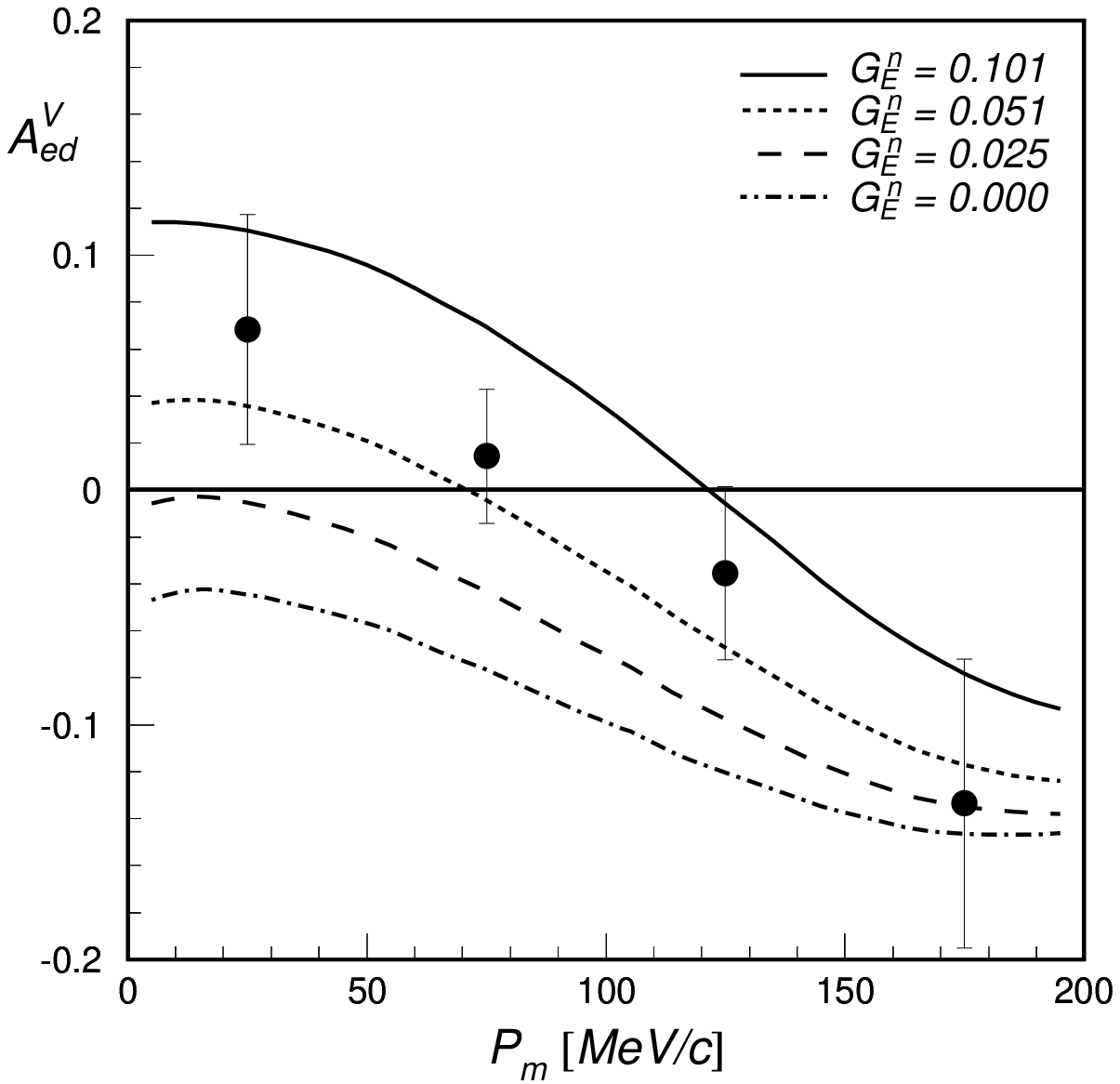}
    \caption{Data and predictions for the sideways asymmetry
      $A^V_{ed}(90^\circ,0^\circ)$ versus missing momentum for the $^2
      \vec{\rm H}(\vec e,e^\prime n)p$ reaction. The curves represent
      the results of the full model calculations of Arenh\"ovel
      \emph{et al.} assuming various values for $G_E^n$.}
    \label{fig3}
  \end{minipage}
  \hspace{\fill}
  \begin{minipage}[t]{75mm}
    \includegraphics[width=75mm]{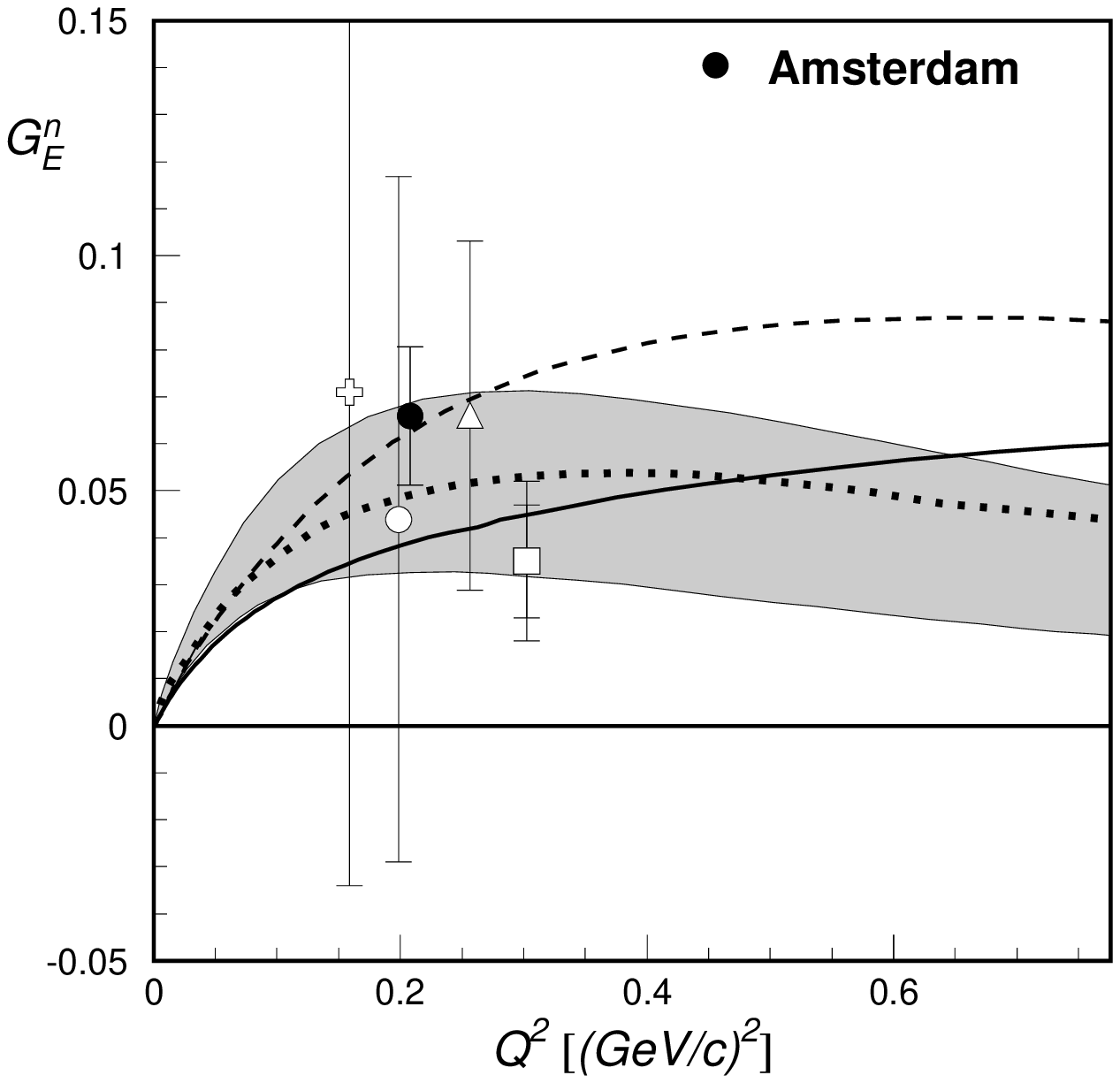}
    \caption{Data and theoretical predictions for $G_E^n$.
      The solid circle shows our result, the cross, open circle, and square 
      the results from double polarization measurments on 
      $^3$He~\protect\cite{genhe} and the triangle on 
      $^2$H~\protect\cite{gendeut}. The shaded area 
      indicates the systematic uncertainty from Ref.~\protect\cite{Platchkov}. 
      The dotted curve shows the results of Ref.~\protect\cite{Galster}, 
      while the solid and dashed curves represent the predictions of the 
      VMD-model of Ref.~\protect\cite{GK}.
      }
    \label{fig4}
  \end{minipage}
\end{figure}

Figure~\ref{fig3} shows the spin-correlation parameter for the $^2\vec
{\rm H}(\vec e,e^\prime n)p$ channel as a function of missing
momentum.  The data are compared to the predictions of the full model
of Arenh\"ovel \emph{et al.}\cite{Arenh}, assuming the dipole
parameterization for the magnetic form factor of the neutron, folded
over the detector acceptance with our Monte Carlo code for various
values of $G_E^n$.  Full model calculations are required for a
reliable extraction of $G_E^n$. We extract $G_E^n(Q^2=0.21 ({\rm
  GeV}/c)^2) = 0.066 \pm 0.015 \pm 0.004$, where the first (second)
error indicates the statistical (systematic) uncertainty.

In Fig.~\ref{fig4} we compare our experimental result to other data
obtained with spin-dependent electron scattering.  The figure also
shows the results from Ref.~\cite{Platchkov}. It is seen that our
result favors their extraction of $G_E^n$ which uses the Nijmegen
potential.

\section{Conclusions}
In summary, we presented the first measurement of the sideways
spin-correlation parameter $A^V_{ed}(90^\circ,0^\circ)$ in quasifree
electron-deuteron scattering from which we extract the neutron charge
form factor at $Q^2 = 0.21$ (GeV/$c$)$^2$. When combined with the
known value and slope \cite{Kop97} at $Q^2 = 0$ (GeV/$c$)$^2$ and the
elastic electron-deuteron scattering data from Ref. \cite{Platchkov},
this result puts strong constraints on $G_E^n$ up to $Q^2 = 0.7$
(GeV/$c$)$^2$.

We would like to thank the NIKHEF and Vrije Universiteit
technical groups for their outstanding support and Prof. H. Arenh\"
ovel for providing the calculations.  This work was supported in part
by the Stichting voor Fundamenteel Onderzoek der Materie (FOM), which
is financially supported by the Nederlandse Organisatie voor
Wetenschappelijk Onderzoek (NWO), the National Science Foundation
under Grants No.  PHY-9504847 (Arizona State Univ.), US Department of
Energy under Grant No. DE-FG02-97ER41025 (Univ. of Virginia) and the
Swiss National Foundation.

\end{document}